\documentclass[mathleft
]{an}
\usepackage{graphicx}
\usepackage{times}
\begin{document}

% The following seven commands are intended for editorial usage and should be ignored by
% the author(s).
\Pagespan{875}{880}% Document's page range. 
% If second parameter is left empty, the last page is computed automatically.
\Yearpublication{2008}%
\Yearsubmission{2008}%
\Month{09/10}%   
\Volume{329}%  
\Issue{9/10}% 
 \DOI{10.1002/asna.200811065}% 
\sloppy

\title{The Gaia Project - technique, performance and status}

\author{S. Jordan\inst{1}\fnmsep\thanks{Corresponding author:
  \email{jordan@ari.uni-heidelberg.de}\newline}
%Example 
%for footnote, note the usage of the \texttt{fnmsep}
%command as separator between institute number and footnote mark} 
}
\titlerunning{The Gaia Project - technique, performance and status}
\authorrunning{S. Jordan}
\institute{
Astronomisches Rechen-Institut, Zentrum f\"ur Astronomie (ZAH), M\"onchhofstr. 12-14, D-69120 Heidelberg, Germany}

\received{August 2008}
\accepted{September 19. 2008hopefully soon}
\publonline{2008}


\abstract{%
Gaia is a satellite mission of the ESA, aiming at absolute astrometric measurements of about one billion stars
(all stars down to $20^{\rm th}$ magnitude, with unprecedented accuracy. Additionally, magnitudes and colors will be obtained for all these stars, while
radial-velocities and spectral properties will be determined only for  bright objects ($V<17.5$). 
At  $15^{\rm th}$ magnitude Gaia aims at an angular accuracy of 20 microarcseconds ($\mu$as). This goal can only be reached if the geometry of the telescopes, the detectors,
and the pointing of Gaia at each moment (``attitude'') can be inferred from the Gaia measurements itself  with $\mu$as accuracy. 
}

\maketitle

\begin{figure}
\includegraphics[width=0.48\textwidth]{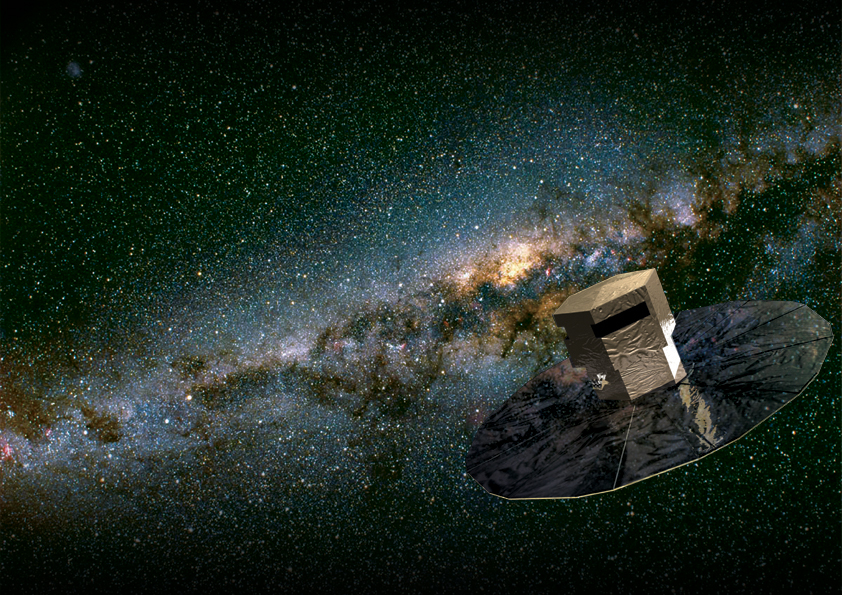}
\caption{Artist's impression of the Gaia satellite scanning the stars of the Milky Way (Image: ESA Medialab)
}
\label{f:gaia_artist_impression}
\end{figure}

\section{Introduction}
The orbit of the Gaia mission has been chosen to be a controlled Lissajous orbit (amplitudes of 360,000 and 100,000 km along and perpendicular to the ecliptic) around the Lagrangian point L2 of the Sun-Earth system (about 1.5 million km in
anti-sun direction) in order to have a quiet environment for the payload in terms of thermo-me\-cha\-nical stability. Another advantage of this position is
the possibility of uninterrupted observations, since the Earth, Moon and Sun all lie inside Gaia's orbit.
The aim of Gaia is to perform absolute astrometry quasi-simultaneously on the whole celestial sphere, rather than differential measurements in a small field of view.

Compared to the precursor mission HIPPARCOS (1989-1993). the much larger telescope of Gaia, the much higher sensitivity of the detectors, the more stable environment at
L2, and other technical progress allows to measure 10,000 times more objects (a factor of 1000 improvement in limiting brightness) and reach a 50 times higher accuracy for 100 times fainter objects.

The basic design of Gaia has been described by  Perryman et al. (2001) and ESA (2000).
However, in 2002 a major reduction in size, complexity, and cost (as well as a degradation
of the astrometric performance by about a factor of two) was made.

More than 300 scientists and computer experts from 18 different countries are working on the Gaia project, not included the work by employees at ESA or the industry. The main
contractor on the industrial side is EADS Astrium.

The main purpose of this paper is to briefly describe the technique of the Gaia project, its performance and status.
More details and up-to-data-information on Gaia can at all time be retrieved from the ESA-RSSD homepage (the URL is provided at the end of the References).

The scientific applications of Gaia will be the topic of 
the contribution by N. Walton in these proceedings.

\section{Gaia's schedule}
The first proposal for a HIPPARCOS successor was submitted to ESA in 1993. It was accepted as a ``Cornerstone Mission'' in
2000. In 2006 the industrial phase began, and in 2007 the Preliminary Design Review was successfully completed.

Gaia is currently scheduled to be launched from Kourou, French Guiana,
in December  2011 with a Soyuz-ST rocket (which includes a restartable Fregat upper stage). Initially
the Fregat-Gaia composite will be placed into a parking orbit, after which a single Fregat boost injects Gaia
on its transfer trajectory towards the L2 Lagrange point. In order to keep Gaia in an orbit around L2, the spacecraft must
perform small maneuvers every month.
After a commissioning phase Gaia will measure the sky for five years with a possible extension for
another year. Subsequently, the final catalog, which includes astrometric and photometric information, near-infrared spectroscopy, 
radial-velocity determinations, and a classification of the objects, will be produced. The completion of the Gaia project
is intended to be around 2020.

It is planned to produce one or more intermediate catalogues in the course of operations. 
The exact dates for such data releases will be carefully decided on to avoid disseminating
insufficiently validated data.

\begin{figure}
\includegraphics[width=0.48\textwidth]{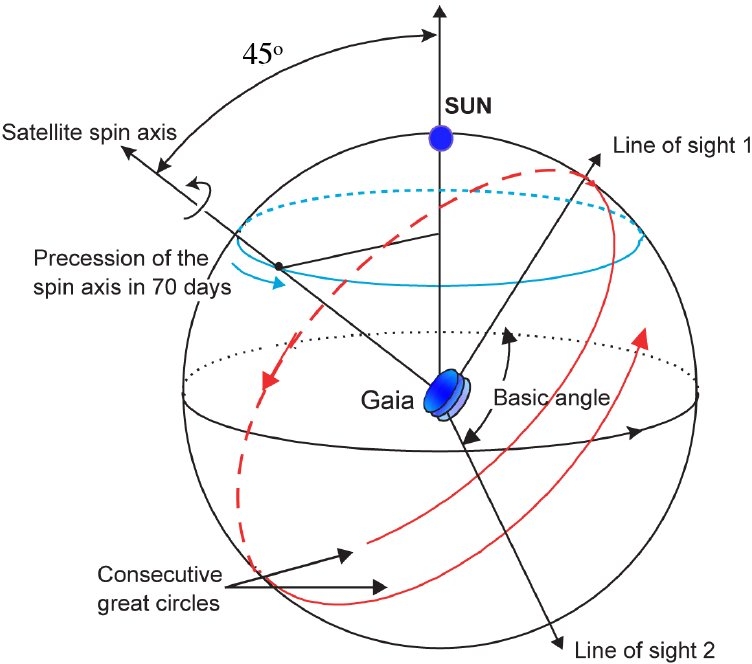}
\caption{Scanning principle of Gaia: The constant spin rate of $60^{\prime\prime}/$s corresponds to one revolution
(great-circle scans) in six hours. The angle between the slowly precessing spin axis and the Sun
 is maintained at an aspect angle of 45$^\circ$ . The basic angle is between the two fields of view is constant at 106.5$^\circ$.
Figure courtesy Karen O'Flaherty, ESA)
}
\label{f:gaia_scanning_principle}
\end{figure}

\section{The measurement principle}
In order to perform high-precision absolute astrometry, Gaia -- like its predecessor Hipparcos -- (see Lindegren 2004)
\begin{itemize}
 \item  simultaneously observes in two fields of view (FoVs) separated by a large
``basic angle'' (106.5$^\circ$), This allows large-angle measurements to be as precise as small-\-scale
ones. In this way it is possible to use the many references stars at large angular separation which have a very different parallax movement, so that
the distance determinations do not suffer from the uncertainties of differential (small-field) angular measurements.
\item  roughly scans along great circles leading to strong ma\-the\-matical closure conditions. This one-dimensionality
also has the advantage that the measured angle between two stars, projected along-scan, is to first order independent of the
orientation of the instrument., 
\item scans the same area of sky many times during the mission
under varying orientations.
\end{itemize}

The angular precision $\sigma$ of a single astrometric one-dimensional measurement is determined by the size of the aperture of each of Gaia's two telescopes
($D=1.4$\,m in along-scan direction), the wavelength $\lambda$ of the incident light from the source,  and the number of photons $N$  reaching the focal plane (which depends on the brightness of the measured
source) during the integration time of the CCD detector:
$$\sigma\approx \frac{\lambda}{D}\frac{1}{\sqrt{N}}.$$

In reality, the real accuracy that can be reached must take into account the spatial sampling of the signal, attitude high-frequency disturbances and other instrumental limitations.
Doing many such measurements for each object at different times and in different directions (position angles on the sky) will allow to derive all five astrometric parameters (two mean coordinates,
two proper motion components, and the parallax), plus additional parameters in the case of binaries.

\section{The nominal scanning law}

\begin{figure}
\includegraphics[width=0.48\textwidth]{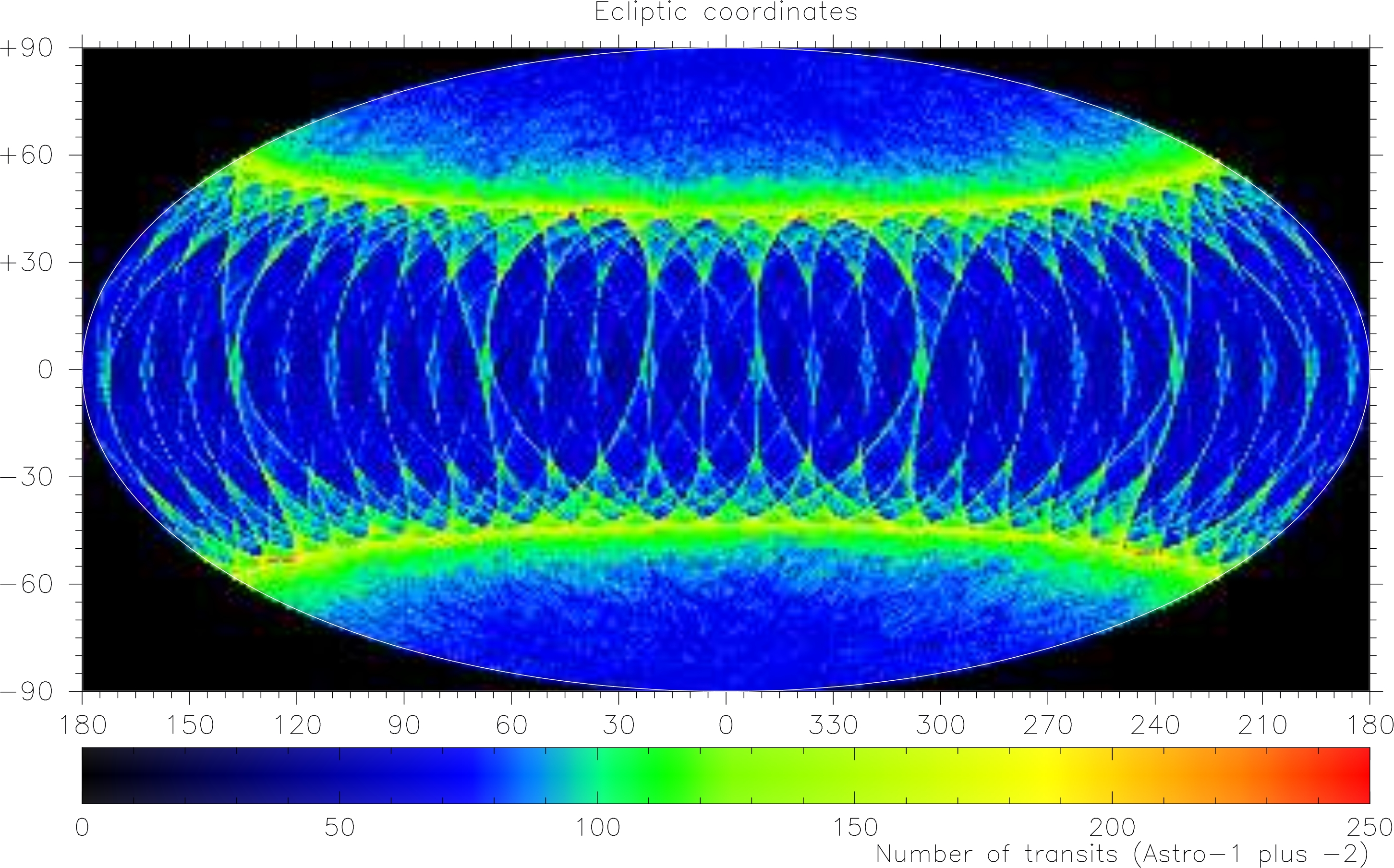}
\caption{During its operational lifetime, Gaia will continuously scan the sky, roughly along great circles, according to a carefully selected pre-defined scanning law. The characteristics of this law, combined with the across-scan dimension of the astrometric fields of view, result in the above pattern for the distribution of the predicted number of transits on the sky in ecliptic coordinates.
 Figure courtesy J. de Bruijne, ESA.
}
\label{f:gaia_scanning_ecliptic}
\end{figure}

\begin{figure}
\includegraphics[width=0.48\textwidth]{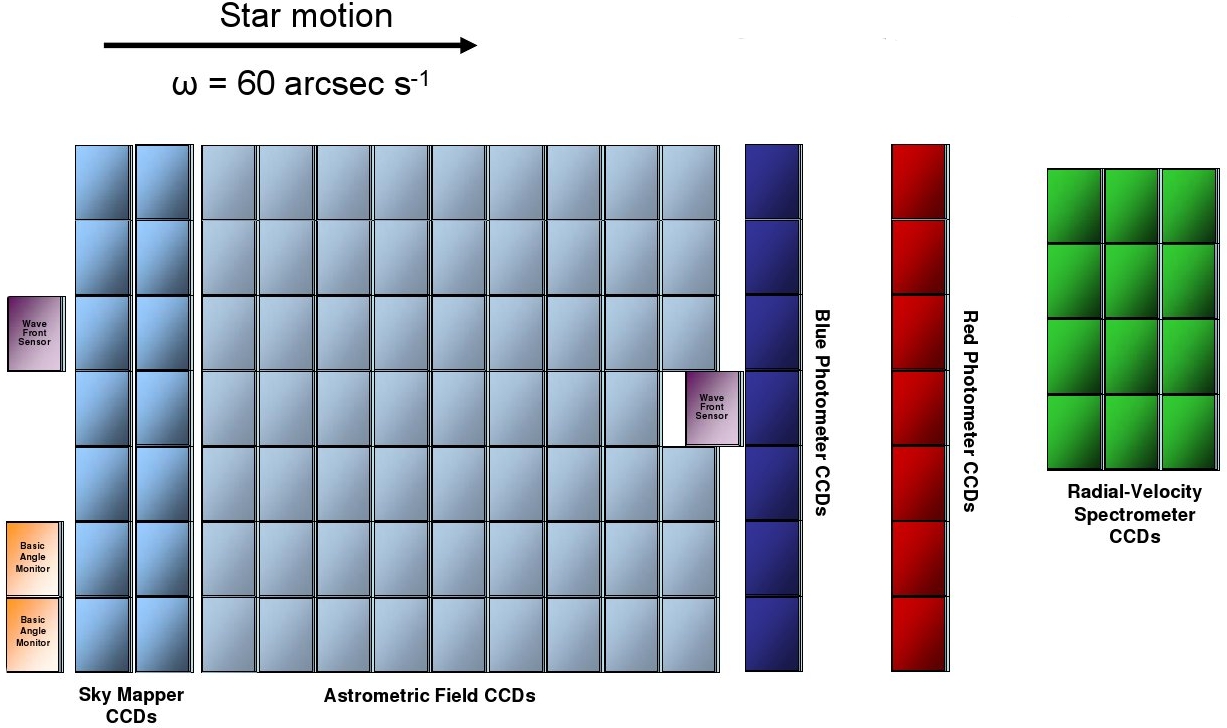}
\caption{Gaia's focal plane. The images of the stars move from left to right. The two CCD strips ($2\times 7$) to the left correspond to the Sky Mapper (SM), adjacent to the right are
the 62 CCDs of the Astrometric Field (AF), followed by the CCDs of the Blue Photometer (BP), Red Photometer (RP), and the Radial-velocity Instrument (RVS). Additionally two CCDs of the
Wavefront Sensor and two for the Basic Angle Monitor are shown. 
 Figure courtesy A. Short \&\ J. de Bruijne, ESA.
}
\label{f:gaia_focal_plane}
\end{figure}

The conditions mentioned in the previous section are fulfilled by Gaia's so-called nominal scanning law (see Fig.\,\ref{f:gaia_scanning_principle}).
The satellite will spin around its axis with a constant rotational period of 6 hours. The spin axis will precess
around the solar direction with a fixed aspect angle of 45$^\circ$ every 63.12 days. A somewhat larger solar-aspect angle would
be optimum, but thermal stability and power requirements limit it to $45^{\circ}$.

On average, each object in the sky is transiting the
focal plane about 70 times  during the 5 year nominal mission duration. Most of the times, an object transiting through
one FoV is measured again after 106.5 or 253.5 minutes (according to the basic angle of 106.5$^\circ$) in the
other FoV. At each transit of the image of of an object, ten single measurements with ten CCDs are performed.

The essential measurements are performed along-scan, the across-scan positions only slightly increase the final precision. Since, due to the nominal scanning law, all 
objects are measured under different scan directions in the course of the mission, the along-scan ones alone can build up a rigid sphere of coordinate measurements.

After about six months every point of the sky is scanned at three or more distinct epochs so that after about 18 months the five astrometric parameters  can be separated from each other. 

\section{The payload}
The Gaia payload consists of two telescopes, a heavy on-board data processing and storage facility, and three scientific instruments mounted on a single optical bench: The astrometric instrument,
the photometers, and a spectrograph to measure radial velocities. 

Gaia has two telescopes, one for each FoV, which consist of one primary mirror of size 1.45\,m$\times$0.5\,m, a secon\-dary and a tertiary  mirror.
The light from both telescopes, which are made up of silicon carbide, is combined to a common focal plane with 106 CCDs by a beam combiner at their exit pupil. An intermediate image is used for field discrimination.
The total focal length is 35\,m. The torus, on which the telescopes and instruments are mounted, requires a thermal stability of some $\mu$K, which is reached by a passive thermal design.  A sunshield of 11\,m diameter, which is partially covered with solar cells to produce electricity, protects against the incident sun light.

The size of the focal plane is 420\,mm$\times$850\,mm  (see Fig.\,\ref{f:gaia_focal_plane}). It consists of 14 Sky Mapper (SM) CCDs, 62 Astrometric Field (AF) CCDs, 7 CCDs for the Blue Photometer (BP),
7 for the Red Photometer (RP), and 12 for the Radial Velocity Spectrometer (RVS). Additionally, two CCDs are used for the Wave Front Sensor and two for the Basic Angle Monitor.

All CCDs operate in the so called Time Delay Integration (TDI) mode, i.e. accumulated charges
of the CCD are transported across the CCD in synchrony with the images.

During a transit through either field of view the image of a star (or an asteroid or quasar) is first registered by one of the SM CCD strips (SM1, if a star passes through FoV 1, SM2 for FoV 2). The SM shall automatically detect all
objects down to 20$^{\rm th}$ magnitude (including variable stars, supernovae, micolensing events, and solar system objects). It also has to reject prompt-particle events (``cosmics''). The on-board
processing of the SM images also
performs the first astrometric measurement in along- and across-scan direction.

In order to additionally filter out false detections, the image of an object must be confirmed on the first strip of the AF in order to be processed any further. If this check is passed, the other AF
CCDs can perform their main goal, the high-precision  measurement of the along-scan  position. This measurement actually consists of the time when the centroid of the image is entering the CCD's read-out
register. Since Gaia has 9 AF strips, every FoV transit delivers 9 such measurements from the AF and 1 from the respective SM. Additionally, the centroid's across-scan position is measured in AF1 and,
-- for calibrational  purposes -- in all other AF strips as well for a small sample of images. 

 In order to reduce the data rate and the read-out noise, only small ``windows''
around each target star, additionally binned in across-scan direction depending on the object's magnitude,
are read out and transmitted to the ground (faint-star windows consist of 6 along-scan TDI pixels, bright stars have an extension of 18). 24 hours of observation data are downlinked  during the 7-8 hours of ground contact per day.
The data rate to the ground station Cebreros in Spain is about five Mbit/s. When Gaia scans along the plane of the Milky Way, the seven hours of contact per day is not enough to download
all data even taking into account the large on-board memory buffer. Therefore, it is planned to increase the downlink capacity with  a second ground station (New Norcia in Australia) during these times of 
galactic plane scans.

The maximum stellar density, important for scans over globular clusters or Baade's window,  where Gaia can perform measurements is 3 million stars down to $20^{\rm th}$ magnitude, but only 600,000 stars per square degree (or one star per six square arcseconds) can actually be measured.  The technical reason for this limitation is that Gaia can read out only five active windows at each moment (TDI beat) on each CCD.
In principle it is possible to temporally activate a mode of reduced precession speed so that more frequent scans over e.g. Baade's windows would be possible. Since the data-loss of stars is random,
more stars have a good  chance to be measured that way. However, it is not decided yet whether such a mode will be actually switched on.

Multi-colour photometry is provided by
two low-reso\-lotion fused-silica prisms dispersing all the
light entering the field of view in the along-scan direction prior to detection.
The Blue Photometer (BP) operates in the wavelength
range 3300--6800\,\AA; the Red Photometer (RP) covers the
wavelength range 6400--10500\,\AA.

The RVS is a near-
infrared (8470 -- 8740\,\AA), medium resolution spectrograph:
$\mathrm{R} = \lambda / \Delta \lambda = 11\,500$. It is illuminated by the same
two telescopes as the astrometric and photometric instruments.

\section{The astrometric solution}
The  astrometric core solution will be based on about  $10^8$ primary stars, which means to solve for
some $5\times 10^8$ astrometric parameters (positions, proper motions, and parallaxes). However, the attitude of the
satellite (parametrised into $\sim 10^8$ attitude parameters over five years) must also only be determined with high
accuracy from the measurements itself. Additionally, a few million calibrational parameters must be solved for to describe the geometry of the
instruments, and finally, deviations from general relativity are accounted for by solving for some
post-Newtonian parameters.

The number of observations for the $10^8$ primary stars is about $6\cdot 10^{10}$.
The condition equations
connecting the unknowns to the observed data are
non-linear but well linearised  at the
sub-arcsec level. Direct solution of the corresponding least-squares
problem is infeasible, because
the large number of unknowns and their strong inter-connec\-tivity
prevents any useful decomposition of the problem into
manageable parts. The proposed method is a block-iterative scheme.
 Intensive tests are currently under way and  have already
demonstrated its feasibility with $10^6$ stars assuming realistic random and systematic
errors in the initial conditions.

The $l^{\rm th}$ time measurement on an SM or AF CCD $t_l$ (the read-out time of the along-scan image centroid) depends on the parameters of the source on the sky $\vec{s}_i$ (position $\alpha, \delta$, proper motion
$\mu_\alpha, \mu_\delta$, and the parallax $\varpi$), the attitude parameters (spline coefficients) $\vec{a}_j$, geometric calibration parameters (e.g. basic angle; position,  distortion, and orientation
of the CCDs) $\vec{c}_k$, global parameters (e.g. the post-Newtonian parameter $\gamma$), and some auxiliary data (ephemeris of solar-system objects, Gaia's orbit, etc) $\vec{A}$ (Lindegren et al. in prep.):
$$t_l=f_{\rm AL}(\vec{s}_i.\vec{a}_j, \vec{c}_k, \vec{g}, \vec{A}) + \rm{noise}$$.

In an analogous way the across-scan position $p_L$ (pixel position of the along-scan centroid) is given by
$$p_l=f_{\rm AC}(\vec{s}_i.\vec{a}_j, \vec{c}_k, \vec{g}, \vec{A}) + \rm{noise}$$.

For Gaia, the light deflection from the Sun is very large all over the sky. In the direction perpendicular to the Sun, the angular change of position is
still 4 milliarcseconds, about 200 times Gaia's accuracy. In the Gaia project even the bending of light due to the major planets cannot be neglected. Moreover, the accuracy
of the Einstein's theory of general relativity to predict the light-deflection can be tested with an accuracy of some $10^{-7}$ (in the PPN parameter $\gamma$).

With provisional values of the source, attitude, calibration, and global parameters, the models for the functional along-scan $f{\rm AC}$ and
across-scan behaviour 
$f_{\rm AC}$ of Gaia allows to predict values for $t_l$ and $p_l$. The difference between the observed and predicted values allows a correction of the
provisional parameter values with a least-squares method.

A linearisation leads to the observation equations
$$t_l^{\rm obs}-t_l^{\rm pred}=
\frac{\partial f_{\rm AL}}{\partial \vec{s}_i}
+\frac{\partial f_{\rm AL}}{\partial \vec{a}_j}
+\frac{\partial f_{\rm AL}}{\partial \vec{c}_k}
+\frac{\partial f_{\rm AL}}{\partial \vec{g}}+\rm{noise}$$

and

$$p_l^{\rm obs}-p_l^{\rm pred}=
\frac{\partial f_{\rm AC}}{\partial \vec{s}_i}
+\frac{\partial f_{\rm AC}}{\partial \vec{a}_j}
+\frac{\partial f_{\rm AC}}{\partial \vec{c}_k}
+\frac{\partial f_{\rm AC}}{\partial \vec{g}}+\rm{noise}.$$

All the parameters are strongly connected to each other, so that it cannot be easily broken into a number of independent solutions.
There is also no simple way to reduce the equations to a smaller structure and size using sparse matrix algebra. Therefore, a direct mathematical
solution seems unfeasible. Note, however, that the strong connectivity is a desired feature that makes the astrometric solution ``rigid'' and accurate.

Since no direct solution is possible, the practical approach is to calculate the 
corrections (updates)  separately for $\Delta \vec{s}$,
$\Delta \vec{a}$, $\Delta \vec{c}$, and $\Delta \vec{g}$, keeping the other values constant (block iteration).
Each partial problem is relatively easy to solve, because the connectivity is given up. 
Cyclically the four partial problems are solved until convergence is reached. Due to the strong correlations
between the four block (particularly the two blocks between attitude and source parameters), a large number of iterations are needed.
However, it has been demonstrated that it is possible to reduce the computing time to a manageable level by convergence acceleration techniques.

This block-iterative scheme is called the Astrometric Global Iterative Solution (AGIS).
This mathematical approach leads to an astrometric ``self-calibration''. In order to reach this calibrational task, the instrument must be very stable over longer time scales.

Later on, with a good solution for the attitude and the geometric parameters based on the
measurements of the  $10^8$ primary stars, the remaining $9\cdot 10^8$ stars can
be linked into the system.
Currently, the AGIS is tested with simulated data (bearing systematic and random errors) for $10^6$ primary sources (which means 1\%\ of the
full problem). A parallel computer cluster with 22 nodes is used on which 45 AGIS iterations take about 70 hours. The system will be successively upgraded to cope with the full problem in about 2012.

In the end, AGIS will determine about 5 billion astrometric parameters, about 150 million unknowns for the attitude, and 10-50 million other calibration parameters from $10^{12}$ individual
astrometric measurements. 

The precision of the astrometric parameters of individual stars
depends on their magnitude and color, and to a lesser
extent on their location in the sky. Sky-averaged values
for the expected parallax precision are displayed in
Table\,\ref{tab:astrom-performance}. The corresponding figures for the
coordinates and for
the annual proper motions are similar but slightly smaller
(by about 15 and 25\%).
Note, that a parallax accuracy of  20 microarcseconds  (the thickness of a human hair seen from a distance of 200 km) for a $15^{\rm th}$ magnitude star
 translates into an accuracy of the distance determination of
0.1\%\ for 50\,pc, and 1\%\ for 500\,pc.

\begin{table}[htb]
\caption{End-of-mission parallax precision in microarcseconds.
  Representative values are shown for unreddened stars of the indicated
  spectral types and V~magnitudes.
  The values are computed
  using the actual Gaia design as input.
  The performance calculation used does not include the effects of radiation
  damage to the CCDs.
  \label{tab:astrom-performance}}
\begin{center}
\begin{tabular}{|c|c|r|}
\hline
Star type & V magnitude   & nominal \\
          &               & performance  \\
\hline
          &  $<$ 10     &   5.2        \\
B1V       &   15        &   20.6       \\
          &   20        &  262.9       \\
\hline
          &   $<$ 10    &   5.1        \\
G2V       &   15        &   19.4       \\
          &   20        &  243.4       \\
\hline
         &  $<$ 10     &   5.2        \\
M6V       &   15       &   8.1        \\
          &   20       &   83.9       \\
\hline
\end{tabular}
\end{center}
\end{table}

\section{What does this accuracy mean?}
In order illustrate picture what an accuracy of 20\,$\mu$as means, consider  the following examples:
\begin{itemize}
\item Proper motion:
\begin{itemize}
 \item $M=+10$: A star with an absolute magnitude of $+10$ has $15^{\rm th}$ at 100\,pc. At this distance, an accuracy of the proper motion of
 20\,$\mu$as/year corresponds to 10\,m/s, i.e. planets can be found around about half a million stars (Jupiter moves the sun by 15 m/s).
 \item $M=0$: Stars with an absolute magnitude of zero could be measured down to 1\,km/s at 10\,kpc (i.e. that even the lowest-velocity stellar populations can be kinematically studied throughout the
entire galaxy). 
\item $M=-3.5$: The motion of bright stars with $M=-3.5$ can be studied down to 5\,km/sec, i.e. that the internal kinematics of the Magellanic Clouds at 50\,kpc can be studied in as much detail as 
the solar neighbourhood can be now (5\,km/s=2.5\,mas/a at 400\,pc).
\item $M=-10$: For the brightest stars velocities of 100 km/s can be still detected at 1\,Mpc, i.e. a handful of very luminous stars in M\,31 will show that galaxy's rotation.
\end{itemize}
\item Parallaxes:
\begin{itemize}
\item 20\,$\mu$as=1\%\ of 0.5\,kpc, i.e. the six-dimensional structure of the Orion complex can be studied with 5\,pc depth resolution.
\item  20\,$\mu$as=10\%\ of 5\,kpc, i.e. a direct high-precision distance determination is possible even for very small stellar groups throughout most of our Galaxy.
,\item  20\,$\mu$as=100\%\ of 50\,kpc, i.e. a direct distance determination of the Magellanic Clouds is at the edge.
\end{itemize}
\item Linear sizes:
\begin{itemize}
\item 20\,$\mu$as=1 solar diameter at 0.5\,kpc, i.e. normal sun\-spots just do not disturb the measurements, but Jupi\-ters do, what makes them
detectable to some extend.
\end{itemize}

\end{itemize}

\section{The spectro-photometric and RV observations}
The most precise photometry for each source will be based on
the unfiltered AF observations with a broad passband between 3500-10,000\,\AA.

Colour information will be available
from the red (RP) and blue (BP) spectrophotometric fields for all sources – these data
will be in the form of mission-averaged low-resolution spectra. 

Although the astrometric instrument contains no refractive optics, the images are
slightly chromatic because of the wavelength dependence of diffraction. Therefore, the
 photometric colour determinations are not only nice to have astrophysically, but are needed to correct for the 
relative displacement between early-type and  very red stars (chromaticity) which may be as large as 1 milliarcsecond.

The photometry provided by Gaia will be unprecedented in homogeneity and depth. The precision of the
brightness determination with the AF  will be between 1 and 10 millimag at $19^{\rm th}$ magnitude.

Moreover, radial velocities are measured with a precision between 1\,km/sec (V=11.5) and
30\,km/sec (V=17.5). 
The measurements of radial velocities are important to correct for perspective acceleration which is induced by the
motion along the line of sight.

The spectroscopic RVS measurements complement the astrometric ones, so that all three spatial and all three velocity components are available, at least for the brighter stars.
By analysing the RVS spectra and the BP/RP photometry, also the atmospheric parameters $T_{\rm eff}$, $\log g$, and the chemical composition can be determined for a large fraction of the 
observed stars.

\section{Radiation-induced CCD damages}
At L2 Gaia is outside the Earth's magnetosphere and fully exposed to high-energy particles (mostly protons) from cosmic rays and the Sun. With its limited
weight budget for a Soyuz-Fregat launch and the large size of the CCD array, only minimal shielding for the detectors is possible. 

The major effect on the CCDs comes from protons from the Sun during solar flares, which may be quite frequent during the Gaia mission, because Gaia will be launched close to the next solar maximum. 
 These protons can collide with the silicon atoms of the CCD and may generate a point defect in the 
crystal lattice. As a result, charge traps are produced which can capture the electrons during the read-out in TDI mode. 
These electrons are released again at a slightly late time. 
By this re-distribution of charges within a recorded image,  the point spread function, or  one-dimensional line spread function (LSF) of a source is distorted and the position of the centroids (which are the main quantities in the astrometric measurements) is strongly shifted with respect to an undamaged CCD.
Since only a small window around each star is downlinked to the ground, the tail of trapped charges released outside the windows reduces the (photometric) flux measurements, additionally to the
astrometric effect.

Ground-based investigations of irradiated CCDs are currently undertaken in order to find out the exact behaviour of the CCDs under various conditions. Of particular importance is that
the effect on the LSF depends on the the history: If another image has passed the same region of the CCD seconds  before, the traps are still partially filled. This interdependence
can strongly be reduced if large extra charges are periodically inserted into the CCDs to fill the traps (e.g. every few seconds). 
Then  fewer empty traps are encountered and the centroid shifts as well as the distortions of the LSFs are minimised. Additionally, only the CCD charge history between two charge injections
need to be taken into account for the correction of the effects.

The hope is that, from the ground-based measurement, the radiation damage effects can be so well understood that a full correction for the centroid shifts and the flux reduction is possible. 
However, the data processing of Gaia
is strongly complicated, since a large number of extra parameters have to be taken into account: The density of the traps at each instance, the magnitude of the
source, the time since charge injection, and the time since previous sources moved through the same CCD column. 

It is very probable, that the values for the astrometric end-of-mission performance are degraded somewhat, but a realistic estimation is not yet possible. Additionally, the limiting 
magnitude for the RVS may be brighter than originally planned.

\section{Computational effort}
The total raw data volume is estimated to be of the order of 100 TB, the total amount of processed and archived data is of the order of 1 PB. Current estimated for the 
computational volume is $1.5\cdot 10^{21}$ floating-point operations, but this may actually be a lower limit and  will probably increase due to the correction tasks for the
radiation-induced CCD damages. If the processing of one star (with typically 1000 measurement per star) 
would need 1 second, 30 years of the data analysis would be needed. However, with massively  distributed computation and the faster computers of the time of Gaia's data analysis, this
task will be feasible.

\section{Conclusion}
Since distance determinations are of such fundamental importance in astronomy, it is clear that Gaia's high-accuracy parallax measurements will  influence  basically
all fields in astronomy. Together with the radial-velocity data, the proper  motion determinations will be of particular importance for our understanding of the
stellar dynamics of the Milky Way. For details on the expected progress, see the following paper by N. Walton.

\acknowledgements
Work on Gaia data processing in Heidelberg  is supported by the
DLR grant 50 QG 0501. The figures are courtesy of EADS Astrium and ESA. The author thanks U. Bastian for
the careful reading and valuable discussions and for a draft of Sect. 7.

%\newpage%%%%%%%%%%%%%%%%%%%%%%%%%%%%%%%%%%%%%%%%%%%%%%%%%%%%%%

\end{document}